\documentstyle{mn}
\input psfig

\newif\ifAMStwofonts

\newcommand{\grad}{\mbox{\boldmath $\nabla$}}
\newcommand{\half}{{\small\frac{1}{2}}}
\newcommand{\Mach}{{\cal M}}
\newcommand{\pdr}{\frac{\partial}{\partial R}}
\newcommand{\tom}{\tilde\omega}
\newcommand{\vcirc}{v_{\rm circ}}
\newcommand{\Omegap}{\Omega_{\rm p}}

\newcommand{\alphan}{\alpha_{\rm n}}
\newcommand{\alphanm}{\alpha_{\rm n-2}}
\newcommand{\bu}{\mbox{\boldmath $u$}}
\newcommand{\hu}{\mbox{\boldmath $\hat u$}}
\newcommand{\bU}{\mbox{\boldmath $U$}}

\newcommand{\bM}{\mbox{\boldmath $M$}}
\newcommand{\bn}{\mbox{\boldmath $n$}}

\newcommand{\bp}{\mbox{\boldmath $p$}}
\newcommand{\atan}{\rm atan}

\ifoldfss
  \ifCUPmtlplainloaded \else
    \NewTextAlphabet{textbfit} {cmbxti10} {}
    \NewTextAlphabet{textbfss} {cmssbx10} {}
    \NewMathAlphabet{mathbfit} {cmbxti10} {} 
    \NewMathAlphabet{mathbfss} {cmssbx10} {} 
  \fi
  \ifAMStwofonts
    \ifCUPmtlplainloaded \else
      \NewSymbolFont{upmath} {eurm10}
      \NewSymbolFont{AMSa} {msam10}
      \NewMathSymbol{\upi}     {0}{upmath}{19}
      \NewMathSymbol{\umu}     {0}{upmath}{16}
      \NewMathSymbol{\upartial}{0}{upmath}{40}
      \NewMathSymbol{\leqslant}{3}{AMSa}{36}
      \NewMathSymbol{\geqslant}{3}{AMSa}{3E}

       \let\le=\leqslant
       \let\ge=\geqslant
    \fi
  \fi
\fi 

\ifnfssone
  \newmathalphabet{\mathit}
  \addtoversion{normal}{\mathit}{cmr}{m}{it}
  \addtoversion{bold}{\mathit}{cmr}{bx}{it}
  \newmathalphabet{\mathbfit} 
  \addtoversion{normal}{\mathbfit}{cmr}{bx}{it}
  \addtoversion{bold}{\mathbfit}{cmr}{bx}{it}
  \newmathalphabet{\mathbfss} 
  \addtoversion{normal}{\mathbfss}{cmss}{bx}{n}
  \addtoversion{bold}{\mathbfss}{cmss}{bx}{n}
  \ifAMStwofonts
    \ifCUPmtlplainloaded \else
      \UseAMStwoboldmath
      \makeatletter
      \new@mathgroup\upmath@group
      \define@mathgroup\mv@normal\upmath@group{eur}{m}{n}
      \define@mathgroup\mv@bold\upmath@group{eur}{b}{n}
      \edef\UPM{\hexnumber\upmath@group}
      \new@mathgroup\amsa@group
      \define@mathgroup\mv@normal\amsa@group{msa}{m}{n}
      \define@mathgroup\mv@bold\amsa@group{msa}{m}{n}
      \edef\AMSa{\hexnumber\amsa@group}
      \makeatother
      \mathchardef\upi="0\UPM19
      \mathchardef\umu="0\UPM16
      \mathchardef\upartial="0\UPM40
      \mathchardef\leqslant="3\AMSa36
      \mathchardef\geqslant="3\AMSa3E

       \let\le=\leqslant
       \let\ge=\geqslant
    \fi
  \fi
\fi 

\ifnfsstwo
  \DeclareMathAlphabet{\mathbfit}{OT1}{cmr}{bx}{it}
  \SetMathAlphabet\mathbfit{bold}{OT1}{cmr}{bx}{it}
  \DeclareMathAlphabet{\mathbfss}{OT1}{cmss}{bx}{n}
  \SetMathAlphabet\mathbfss{bold}{OT1}{cmss}{bx}{n}
  \ifAMStwofonts
    \ifCUPmtlplainloaded \else
      \DeclareSymbolFont{UPM}{U}{eur}{m}{n}
      \SetSymbolFont{UPM}{bold}{U}{eur}{b}{n}
      \DeclareSymbolFont{AMSa}{U}{msa}{m}{n}
      \DeclareMathSymbol{\upi}{0}{UPM}{"19}
      \DeclareMathSymbol{\umu}{0}{UPM}{"16}
      \DeclareMathSymbol{\upartial}{0}{UPM}{"40}
      \DeclareMathSymbol{\leqslant}{3}{AMSa}{"36}
      \DeclareMathSymbol{\geqslant}{3}{AMSa}{"3E}

       \let\le=\leqslant
       \let\ge=\geqslant
    \fi
  \fi
\fi 

\ifCUPmtlplainloaded \else
  \ifAMStwofonts \else 
    \def\upi{\pi}
    \def\umu{\mu}
    \def\upartial{\partial}
  \fi
\fi

\title{The Paradox of the Scale-Free Disks}
\author[Jeremy Goodman and N. Wyn Evans]
       {Jeremy Goodman$^1$ and N. Wyn Evans$^2$ \\
        $^1$ Princeton University Observatory, Princeton, NJ
        08544--1001\\
        $^2$ Theoretical Physics, 1 Keble Rd, Oxford, OX1 3NP}

\begin{document}

\maketitle

\label{firstpage}

\begin{abstract}
Scale-free disks have no preferred length or time scale.  The question
has been raised whether such disks have a continuum of unstable linear
modes or perhaps no unstable modes at all. We resolve this paradox by
analysing the particular case of a gaseous, isentropic disk with a
completely flat rotation curve (the Mestel disk) exactly. The heart of
the matter is this: what are the correct boundary conditions to impose
at the origin or central cusp? We argue that the linear stability
problem is ill-posed.  From any finite radius, waves reach the origin
after finite time but with logarithmically divergent phase.
Instabilities exist, but their pattern speeds depend upon an
undetermined phase with which waves are reflected from the origin.
For any definite choice of this phase, there is an infinite but
discrete set of growing modes.  Similar ambiguities may afflict
general disk models with power-law central cusps.  The ratio of growth
rate to pattern speed, however, is independent of the central phase.
This ratio is derived in closed form for non self-gravitating normal
modes.  The ratio for self-gravitating normal modes is found
numerically by solving recurrence relations in Mellin-transform space.
\end{abstract}


\section{Introduction}

A scale-free disk, whether gaseous or stellar, has no characteristic
frequency. It is therefore puzzling how such a disk can distinguish a
characteristic frequency for its marginally stable modes. This paradox
was first pointed out by Zang (1976) in a famous Massachusetts
Institute of Technology Ph. D. thesis (supervised by Alar Toomre),
which explored the stability of the scale-free stellar dynamical disks
with completely flat rotation curves. Zang showed that if a scale-free
disk admits one unstable mode, then a two-dimensional continuum can be
constructed by self-similar scaling. Thinking this improbable, Zang
argued that it is more likely that scale-free disks admit no normal
modes whatsoever.  If true, this is astonishing, as it must hold good
even for the completely cold disk.  Subsequently, the problem was
re-visited by Lynden-Bell \& Lemos (1993), who took the opposite view
and argued that all the modes become unstable together. The
possibility that an entire continuum of modes, each with a different
frequency, might all become simultaneously unstable had earlier been
envisaged by Birkhoff in his classic 1960 textbook on {\it
Hydrodynamics}. If true, this also is astonishing, as the disks would
become unstable at all scales to normal modes of all frequencies.

This paper resolves the paradox by showing how to construct an
infinite but discrete set of normal modes. Section 2 introduces the
disks under scrutiny and sets up the mode equations for the perturbed
quantities. Section 3 examines the easier but still ambiguous problem
of modes without self-gravity, whereas Section 4 deals with the fully
self-gravitating case.

\section{Equilibria and Perturbations}

\subsection{Equilibria}

Our disks are scale-free in every respect.  For definiteness, we
suppose that they are razor-thin and fully self-gravitating in
equilibrium. In fact, these properties are inessential as long as
there is no preferred scale.  Wherever possible, we use the same
notation as in the recent stability analysis of the stellar dynamical
power-law disks carried out by Evans \& Read (1998a,b).  Some basic
formulae are repeated here for convenience and completeness.

In equilibrium, the surface density and self-consistent potential are
\begin{eqnarray}
\Sigma_{\rm eq}(R)&=& \Sigma_0\left(\frac{R}{R_0}\right)^{-\beta-1},
\nonumber\\
\psi_{\rm eq}(R)&=& -\frac{v_\beta^2}{\beta}
\left(\frac{R}{R_0}\right)^{-\beta},
\label{eq:equil}
\end{eqnarray}
in which $v_\beta$ is a constant reference velocity:
\begin{equation}
v_\beta^2= 2\pi G\Sigma_0 R_0\frac{\Gamma\left[\half(1-\beta)\right]
\Gamma\left[\half(2+\beta)\right]}{\Gamma\left[\half(1+\beta)\right]
\Gamma\left[\half(2-\beta)\right]}.
\label{eq:vbeta}
\end{equation}
Here, $\psi$ is a positive quantity; the gravitational acceleration is
$\grad\psi$.  As $\beta\to 0$, $v_\beta^2\to 2\pi G\Sigma_0 R_0\equiv
v_0^2$, and the potential becomes that of the Mestel
disk~\cite{Mestel},
\begin{equation}
\psi_{\rm eq}(R)= -v_0^2\ln\left(\frac{R}{R_0}\right).
\label{eq:mestelphi}
\end{equation}
The reference radius $R_0$ and surface density $\Sigma_0$ appear for
the sake of dimensional consistency but do not introduce preferred
scales, since the replacements
\begin{equation}
R_0\to s\bar R_0,\qquad
\Sigma_0\to s^{-\beta-1}\bar\Sigma_0,\qquad
v_\beta^2 \to s^{-\beta}\bar v_\beta^2,
\end{equation}
preserve the form of equations (\ref{eq:equil})-(\ref{eq:vbeta})
[except in the case $\beta=0$, where the potential
(\ref{eq:mestelphi}) gains a physically unimportant additive
constant].

The disks studied here are gaseous and isentropic, unlike their
stellar dynamical counterparts explored by Evans \& Read
(1998a,b). The axisymmetric and the neutral modes of the gaseous
power-law disks have already been the subject of recent attention
(e.g., Schmitz \& Ebert 1987; Lemos, Kalnajs \& Lynden-Bell 1991; Syer
\& Tremaine 1996).  The square of the sound speed varies with radius
as
\begin{equation}
c^2= \frac{c_\beta^2}{\beta+1}\left(\frac{R}{R_0}\right)^{-\beta}.
\label{eq:csound}
\end{equation}
For convenience, let us introduce a dimensionless ``temperature''
\begin{equation}
\Theta\equiv \frac{c^2}{|d\psi_{\rm eq}/d\ln R|}.
\label{eq:Thetadef}
\end{equation}
Radial hydrostatic equilibrium requires the square of the circular
velocity to be
\begin{equation}
\vcirc^2 = (1-\Theta)v_\beta^2\left(\frac{R}{R_0}\right)^{-\beta}.
\label{eq:vcirc}
\end{equation}
The Mach number is defined as:
\begin{equation}
\Mach \equiv \frac{\vcirc}{c} = \sqrt{\frac{1-\Theta}{\Theta}}.
\label{eq:Mach}
\end{equation}
The angular velocity and epicyclic frequency,
\begin{equation}
\Omega \equiv \frac{\vcirc}{R},\qquad
\kappa \equiv \left[\frac{1}{R^3}\frac{d}{dR}(R\vcirc)^2\right]^{1/2},
\label{eq:freqs}
\end{equation}
both scale as $R^{-\beta/2 -1}$, and their ratio is
\begin{equation}
\frac{\kappa}{\Omega}= \sqrt{2-\beta}.
\label{eq:freqrat}
\end{equation}
We assume $-1\le \beta <1$ so that $\kappa$ is always real.  Toomre's
condition $Q\ge 1$ for local axisymmetric stability~\cite{Toomre64}
is related to $\Theta$ by
\begin{equation}
Q\equiv \frac{c\kappa}{\pi G\Sigma}= 2
\left(\frac{v_\beta^2}{2\pi G\Sigma_0R_0}\right)
\sqrt{(2-\beta)\Theta(1-\Theta)}.
\label{eq:Q}
\end{equation}
In terms of the Mach number, this becomes:
\begin{equation}
Q = 2\sqrt{2-\beta} \left(\frac{v_\beta^2}{2\pi G\Sigma_0R_0}\right)
\frac{\Mach}{1 + \Mach^2}.
\label{eq:QintermsofM}
\end{equation}
The factor in parentheses, which can be read off from equation
(\ref{eq:vbeta}), reduces to unity in the Mestel disk ($\beta=0$).

\subsection{Perturbations}

We consider time-dependent first-order Eulerian perturbations of the
equilibria described above and adopt cylindrical polar coordinates
$(R,\theta, z)$.  Thus,
\begin{eqnarray*}
\Sigma &\to& \Sigma_{\rm eq}(R)~+~\delta\Sigma(R,\theta,t),\\
\psi   &\to& \psi_{\rm eq}(R)~+~\delta\psi(R,\theta,z,t),\\
v_R    &\to& 0~~~~~~~~+~\delta v_R(R,\theta,t),\\
v_\theta &\to& v_{\rm circ}~~~~+~\delta v_\theta(R,\theta,t).
\end{eqnarray*}
Strictly speaking, $\psi_{\rm eq}$ depends upon $z$ as well as $R$,
but we have no occasion to evaluate it outside the plane $z=0$.  As
the equilibrium is independent of angle $\theta$ and time $t$, the
stability analysis is simplified by assuming that the perturbed
quantities have the $(\theta,t)$ dependence characteristic of normal
modes
\begin{equation}
\delta\propto e^{im\theta-i\omega t}.
\label{eq:exfactor}
\end{equation}
Henceforth, we take this factor as read and often write, for example,
$\delta\Sigma(R)$ rather than $\delta\Sigma(R,\theta,t)$.

The linearized dynamical equations are [cf. Binney \& Tremaine, 1987,
chap. 5]
\begin{eqnarray}
-i\tom\delta v_R - 2\Omega\delta v_\theta &=&\pdr\left(\delta\psi
-c^2\frac{\delta\Sigma}{\Sigma}\right),\nonumber\\
-i\tom\delta v_\theta + \frac{\kappa^2}{2\Omega}\delta v_R &=&
\frac{im}{R}\left(\delta\psi
-c^2\frac{\delta\Sigma}{\Sigma}\right),\nonumber\\
-i\tom\frac{\delta\Sigma}{\Sigma} -\beta\frac{\delta v_R}{R} &=&
-\pdr\delta v_R -\frac{im}{R}\delta v_\theta,
\label{eq:lindyn}
\end{eqnarray}
together with Poisson's equation
\begin{equation}
\nabla^2\delta\psi = 4\pi G\delta \Sigma\delta(z).
\label{eq:Poisson}
\end{equation}
We have used the abbreviation
\begin{equation}
\tom\equiv\omega -m\Omega
\label{eq:tom}
\end{equation}
for the Doppler-shifted frequency in the local rest frame of the gas.
Taking the curl of the first two of equations (\ref{eq:lindyn}) and
using the third to eliminate the derivative of $\delta v_R$ produces
\begin{eqnarray}
&&i\tom\left[\frac{1}{R}\pdr(R \delta v_\theta)-\frac{im}{R}\delta v_R
-\frac{\kappa^2}{2\Omega}\frac{\delta\Sigma}{\Sigma}\right] \nonumber
\\ &&~~~~~~~~~~~~~~~~~~~~~~~=\delta
v_R\Sigma\frac{d}{dR}\left(\frac{\kappa^2}{2\Omega\Sigma}\right).
\label{eq:dpvort}
\end{eqnarray}
The quantity
\begin{equation}
\zeta_{\rm eq}\equiv\frac{\kappa^2}{2\Omega\Sigma}
\label{eq:pvort}
\end{equation}
is the equilibrium distribution of potential vorticity (or the curl of
the velocity divided by surface density).  Equation (\ref{eq:dpvort})
states that fluid elements conserve their potential vorticity in a
Lagrangian sense. This is easy to understand physically, as the fluid
elements must preserve both their mass and their circulation (Kelvin's
theorem).

We expect a continuum of modes involving a net change in potential
vorticity.  Such modes will be discontinuous in $\delta v_\theta$ at
the corotation radius where $\tom=0$. This can be anywhere in the
disk.  The eigenfrequencies $\{\omega\}$ are continuously distributed.
We are not interested in these modes because they can never be
unstable ($\omega$ must be real).  If any such mode were to grow or
decay, the net potential vorticity of the disk would have to change,
which is not permitted by our inviscid equations.  The issue of
interest to us is whether there exist growing, non-vortical modes:
that is, disturbances with the $(\theta,t)$ dependence
(\ref{eq:exfactor}) and vanishing total potential vorticity.  The
equations governing the non-vortical modes simplify in the Mestel disk
because $\zeta_{\rm eq}\propto R^{\beta/2}$, which is constant when
$\beta=0$ so that equation (\ref{eq:dpvort}) reduces to
\begin{equation}
\frac{1}{R}\pdr(R \delta v_\theta)-\frac{im}{R}\delta v_R
-\frac{\kappa^2}{2\Omega}\frac{\delta\Sigma}{\Sigma}=0.
\label{eq:curleq}
\end{equation}
Henceforth, we restrict our attention to the Mestel disk.

\section{Modes without Self-Gravity}

In this section, on top of the restriction to $\beta=0$, we neglect
the perturbation in the gravitational potential: that is, we force
$\delta\psi\to 0$.  This is called the ``Cowling approximation'' in
the context of stellar oscillations, where it is justified by the
central concentration of the star's mass or by the short wavelength of
the modes of interest.  Similar justifications could be offered for
the disk modes in the limit $\beta\approx 1$, or for general $\beta$
in the limit of large $m$.  But we invoke neither limit, because we
make no pretense of quantitative accuracy until Section 4.  The true
justification for neglecting the gravitational perturbation is that we
thereby obtain equations that can be solved exactly.  This simplified
problem retains the elements that make the stability problem ambiguous
in the self-gravitating case.

\subsection{A Second-Order Differential Equation}

With $\delta\psi$ neglected, the second of equations (\ref{eq:lindyn})
and equation (\ref{eq:curleq}) can be solved for $\delta v_R$ and
$\delta\Sigma$ in terms of the variable
\begin{equation}
y\equiv R\delta v_\theta
\label{eq:ydef}
\end{equation}
and its radial derivative:
\begin{eqnarray}
\delta v_R &=& -\frac{i}{D}\left(mc^2\frac{\partial y}{\partial R}
- \tom\vcirc y\right),\nonumber\\
\frac{\delta\Sigma}{\Sigma} &=& \frac{1}{D}
\left(\vcirc\frac{\partial y}{\partial R} + m\tom y\right),
\label{eq:sols}
\end{eqnarray}
where the denominator
\begin{equation}
D\equiv \vcirc^2+m^2c^2
\label{eq:Ddef}
\end{equation}
is constant.  Eliminating $\delta v_R$ and $\delta \Sigma$ from the
third of equations (\ref{eq:lindyn}) produces a second-order
differential equation for $y$, which we write in the equivalent forms
\begin{eqnarray}
\frac{\partial^2y}{\partial R^2} + \left(\frac{\tom^2-\kappa^2}{c^2}
-\frac{m^2}{R^2}\right)y &=&0,\nonumber\\
\frac{\partial^2y}{\partial R^2} + \left[
\frac{(\omega-m\Omega)^2-2\Omega^2}{c^2}-
\frac{m^2}{R^2}\right]y &=&0.
\label{eq:yeqn}
\end{eqnarray}

Equation (\ref{eq:yeqn}) is non-singular. In the general $\beta\ne 0$
case, second-order equations can also be obtained for the various
fluid variables, but at least one coefficient will be singular at
corotation ($\tom\to 0$) because of the non-zero potential vorticity
gradient.  In all other respects, equation (\ref{eq:yeqn}) is typical
of the general case.  The wave is evanescent between the Lindblad
resonances, $\tom=\pm\kappa$ or $\omega= m\Omega \pm\kappa$.  In fact,
the WKBJ radial wavenumber becomes real somewhat beyond the Lindblad
resonances (further from corotation) because of the $\tom^2/c^2$ term.
If we had allowed for the self-gravity of the mode, propagation would be
possible between the Lindblad resonances but not in the immediate
vicinity of corotation unless $Q<1$ (see, for example, Binney \&
Tremaine 1987 or Shu 1992)

Equation (\ref{eq:yeqn}) can be reduced to a confluent hypergeometric
function.  Let us make the following transformations of the
independent and dependent variables:
\begin{eqnarray}
z &\equiv& -\frac{2i\omega}{c}R,\nonumber\\
y &\equiv& R^{\half +i\mu}e^{i\omega R/c} w (z)\nonumber\\
  &\propto& z^{\half +i\mu} e^{-z/2} w(z),
\label{eq:transform}
\end{eqnarray}
where $\mu$ is the following dimensionless function of $m$ and the Mach
number (\ref{eq:Mach}):
\begin{equation}
\mu\equiv \half\left[4(m^2-2)\Mach^2-4m^2-1\right]^{1/2}.
\label{eq:mudef}
\end{equation}
There should be no confusion between the dimensionless variable
$z$ and the original cylindrical polar coordinate of the same name.
Equation (\ref{eq:yeqn}) becomes
\begin{equation}
z\frac{d^2 w}{dz^2} + (1+2i\mu-z)\frac{dw}{dz}
-\left(i\mu+im\Mach+\half\right)w =0,
\label{eq:Kumeq}
\end{equation}
which is Kummer's equation~\cite{AS}.  The solution regular at $z=0$
is Kummer's function $M(\half+i\mu+im\Mach,1+2i\mu,z)$.  In fact, if
one uses $-\mu$ instead of $\mu$ in the transformation
(\ref{eq:transform}) of the dependent variable, one obtains an
equation identical to (\ref{eq:Kumeq}) except in the sign of $\mu$,
whose regular solution is $M(\half-i\mu+im\Mach,1-2i\mu,z)$.  Hence,
the two independent solutions of equation (\ref{eq:yeqn}) are
\begin{equation}
y_\pm(z) \equiv z^{\half\pm i\mu}e^{-z/2}
M\left(\half\pm i\mu+im\Mach,1\pm 2i\mu,z\right).
\label{eq:ypm}
\end{equation}
Let us note that Drury (1980) already derived a differential equation
equivalent to (\ref{eq:yeqn}) and used it to find the reflection and
transmission coefficients for the Mestel disk. He did not, however,
construct the normal modes, which we now proceed to do.

\subsection{Boundary Conditions}

The desired eigenfunction is a linear combination of the independent
solutions (\ref{eq:ypm}) that satisfies appropriate boundary
conditions at large and small $R$.  The boundary condition at large
radius is straightforward. We insist on purely outgoing disturbances,
otherwise it is all too easy to manufacture counterfeit instabilities
whose exponential ``growth'' is due solely to increasingly powerful
transmissions from sources at $R=\infty$.  One can see directly from
the second form of equation (\ref{eq:yeqn}) that the radial wavenumber
$k_R\to\pm\omega/c$ as $R\to\infty$ for any value of $m$ provided only
that $\mbox{Real}(\omega)\ne0$.  The outgoing wave is proportional to
\begin{displaymath}
e^{+i\omega R/c}\propto e^{-z/2},
\end{displaymath}
and the signs of the exponents are reversed for the ingoing wave.
Using standard results~\cite{AS}, one sees that the asymptotic
behavior of $y_\pm(z)$ as $|z|\to\infty$ in the cut plane
$-\half\pi<\arg(z)<{\textstyle\frac{3}{2}}\pi$ is
\begin{eqnarray}
&&y_\pm(z)\sim\Gamma(1\pm 2i\mu)\Bigl[\frac{e^{i\pi(\half+im\Mach\pm i\mu)}
z^{-im\Mach}e^{-z/2}}
{\Gamma(\half\pm i\mu -im\Mach)}\nonumber \\
&&~~~~~~~~~~~~~~~~~~~~~~~~~~+~\frac{z^{im\Mach}e^{+z/2}}{\Gamma(\half\pm
i\mu+im\Mach)} \Bigr].
\label{eq:bigz}
\end{eqnarray}
We have to cancel the term in $e^{+z/2}$, so the desired linear
combination is
\begin{equation}
y(z)=\frac{\Gamma(1-2i\mu) y_+(z)}{\Gamma(\half-i\mu+im\Mach)}
- \frac{\Gamma(1+2i\mu) y_-(z)}{\Gamma(\half+i\mu+im\Mach)},
\label{eq:comb}
\end{equation}
or any fixed multiple thereof.

The boundary condition at the origin requires more thought.  First, it
is important to realize that an ingoing wave reaches the origin --
where there is a density cusp -- in finite time. Propagation to the
very centre requires $\Mach>m/\sqrt{m^2-2}$, or equivalently
$0\le\Theta<(m^2-2)/(2m^2+2)$.  This is never possible for $m=1$
disturbances when self-gravity is neglected.
but can happen in sufficiently cold disks for all other
azimuthal wavenumbers. The WKBJ dispersion relation implied by
equation (\ref{eq:yeqn}) is
\begin{equation}
k_R^2 = \frac{(\omega-m\Omega)^2-2\Omega^2}{c^2}-
\frac{m^2}{R^2}.
\label{eq:WKB}
\end{equation}
At sufficiently small radius, where $\Omega(R)\gg |\omega|$, this
reduces to
\begin{equation}
k_R^2\approx\frac{(m^2-2)\Mach^2-m^2}{R^2}~-~\frac{2m\Mach}{Rc}\omega
~+O(\omega^2).
\end{equation}
The numerator of the first term on the right is almost $4\mu^2$
[cf. eq.~(\ref{eq:mudef})]. A comparison with the exact solutions
(\ref{eq:ypm}) shows indeed that it should be $4\mu^2$.
Asymptotically, therefore, the group velocity is
\begin{equation}
V_{\rm g}\equiv\frac{\partial\omega}{\partial k_R}\approx
\mp\frac{2\mu}{m\Mach} c,
\label{eq:Vgroup}
\end{equation}
the upper signs applies if $k_R>0$, and we have assumed $m\ge
2$. Thus, the propagation speed is of order the sound speed,
and an ingoing wave can reach the centre in finite time.

Allowance must therefore be made for a reflected wave.  By analogy
with the situation at large $R$, we require that {\it the origin must
neither absorb nor emit wave energy}.  We proceed to implement this
principle.

At sufficiently small $R$, the Kummer functions $M(\half\pm
i\mu+im\Mach,1\pm 2i\mu,z)\to 1$, and we can read off the ingoing and
outgoing parts of the general solution (\ref{eq:comb}) that is
compatible with the large-$R$ boundary condition:
\begin{eqnarray}
y_{\rm in}&\sim&
-\frac{\Gamma(1+2i\mu)}{\Gamma(\half+i\mu+im\Mach)}
\left(-\frac{2i\omega}{c}R\right)^{-i\mu +\half},\nonumber\\
y_{\rm out}&\sim& \frac{\Gamma(1-2i\mu)}{\Gamma(\half-i\mu+im\Mach)}
\left(-\frac{2i\omega}{c}R\right)^{i\mu + \half}.
\label{eq:yinout}
\end{eqnarray}
(In this paper, the symbol ``$\sim$'' means ``asymptotically approaches''.)
Let us define the flux as the total power crossing a cylinder centred
on the rotation axis.  Then, the fluxes carried by $y_{\rm in}$ and
$y_{\rm out}$ are
\begin{eqnarray}
&&F_{\rm in}= C_{m,\omega}(R)|y_{\rm in}|^2~~\mbox{and}~~
F_{\rm out}=-C_{m,\omega}(R)|y_{\rm out}|^2,\nonumber\\
&&C_{m,\omega}(R)\equiv
\frac{\mu\Sigma(R)\mbox{Real}(\omega)}{\Mach^2+m^2},
\end{eqnarray}
in the limit $\Omega\gg|\omega|$.  The corresponding angular momentum
fluxes can be obtained by dividing by the pattern speed
$\Omegap = \mbox{Real}(\omega)/m$.  The coefficient $C_{m,\omega}(R)$ can be
derived by adapting results from Appendix A of Narayan, Goldreich \&
Goodman (1987), but for our purposes, all that matters is that the
ratio of these fluxes is
\begin{displaymath}
\frac{F_{\rm in}}{F_{\rm out}}=-\frac{|y_{\rm in}|^2}{|y_{\rm out}|^2},
\end{displaymath}
whence we derive the boundary condition
\begin{equation}
\frac{|y_{\rm in}|^2}{|y_{\rm out}|^2}\to 1~~\mbox{as}~R\to 0.
\label{eq:innerbc}
\end{equation}

\subsection{Dispersion Relation and Growth Rates}

The ratio of the moduli of ingoing and outgoing parts of the eigenfunction
(\ref{eq:yinout}) is
\begin{equation}\label{eq:modratio}
\left|\frac{y_{\rm out}}{y_{\rm in}}\right|= \exp[\mu\pi-2\mu\arg(\omega)]
\left|\frac{\Gamma\left(\half+im\Mach+i\mu\right)}
{\Gamma\left(\half+im\Mach-i\mu\right)}\right|.
\end{equation}
Therefore, the inner boundary condition (\ref{eq:innerbc}) requires
\begin{eqnarray}
\arg(\omega)&\equiv&\tan^{-1}\left(\frac{s}{m\Omegap}\right) \nonumber\\
&=& \frac{\pi}{2}-\frac{1}{4\mu}\ln\left|\frac{\cosh\pi(m\Mach+\mu)}
{\cosh\pi(m\Mach-\mu)}\right|.
\label{eq:disper}
\end{eqnarray}
Here, the complex eigenfrequency $\omega$ has been written as $m\Omegap
+ is$, where $s$ is the growth rate and $\Omegap$ is the pattern speed
of the mode.  We have used some identities~\cite{AS} to reduce moduli
of complex gamma functions to elementary functions in the dispersion
relation (\ref{eq:disper}).
\begin{figure}
\psfig{figure=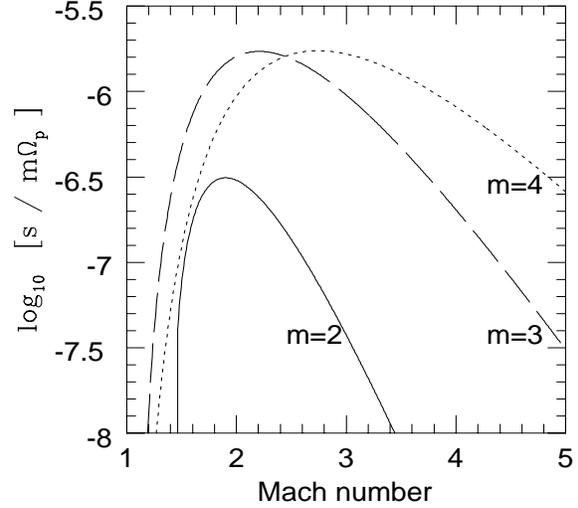,height=8.0truecm,width=8.4truecm}

\caption{Plot of the logarithm of the ratio of growth rate to pattern
speed $s/m\Omegap$ as a function of the Mach number $\Mach$ for non
self-gravitating modes. The diagram shows the dispersion relation for
modes with azithumal wavenumber $m =2,3$ and 4.}
\end{figure}
Fig.~1 shows the variation of the ratio of growth rate to pattern
speed with the hotness of the disks for various azimuthal
wavenumbers. For any temperature or Mach number, only this ratio is
fixed by the dispersion relation. The actual values of $s$ or
$\Omegap$ are indeterminate. This is because there is insufficient
physical information to deduce the phase shift of waves reflected from
the origin.

It may help to give an example of the sort of inner boundary condition
that {\it would} fix the required phase and thereby yield a definite
discrete spectrum of modes.  Suppose the origin is enclosed by a rigid
wall of small radius $R_0\ll |v_{\rm circ}/\omega|$.  Then since no
gas penetrates the wall from outside, $\delta v_R=0$ at $R=R_0$.
Using eq.~(\ref{eq:sols}) to write $\delta v_R$ in terms of $y=y_{\rm
in}+y_{\rm out}$, and using eq.~(\ref{eq:yinout}) to estimate $\partial
y/\partial R$, one finds the boundary condition
\begin{equation}
\frac{y_{\rm out}(R_0)}{y_{\rm in}(R_0)}=\frac{\half+m\Mach-i\mu}
{\half+m\Mach+i\mu}\equiv e^{i\phi_0}
\end{equation}
to leading order in $R_0\ll\vcirc/\omega$.  Notice that the
intermediate expression above indeed has unit modulus, so that
$\phi_0$ is real: this is a consequence of the fact that a rigid wall
absorbs no mechanical energy.  On the other hand, the asymptotic
expressions (\ref{eq:yinout}) yield
\begin{equation}
\arg\left(\frac{y_{\rm out}}{y_{\rm in}}\right)= F(m,\Mach)
~+2\mu\ln\left|\frac{2\omega R}{c}\right|,
\end{equation}
where $F$ is independent of $R$ and $\omega$.  Comparing these last
two equations, we find that
\begin{equation}
\ln|\omega|=\frac{\phi_0+2\pi n -F(m,\Mach)}{2\mu}~-\ln R_0,
\end{equation}
where $n$ can be any integer.  Thus the eigenfrequencies form a
discrete geometric progression: $\omega_{n+1}\sim
e^{\pi/\mu}\omega_n$.  The actual frequencies depend upon the choice
of boundary radius $R_0$, although the ratio of their real to
imaginary parts does not.  For a different kind of physical boundary
condition (perhaps a free inner edge, for example), one would have a
different $\phi_0$ and hence a different correspondence between $R_0$
and the eigenfrequencies.  However, $\arg(\omega_n)$ is always given
by eq.~(\ref{eq:disper}), provided $\omega_n\ll\vcirc/R_0$.

\section{Modes with Self-Gravity}

Let us now return to the fully self-gravitating case.  In Section 4.1,
we derive a recurrence relation between the Mellin transforms of the
perturbed fluid quantities. This relates Mellin transforms with
arguments with the same real part but differing by integers in their
imaginary part. The boundary conditions for this recurrence relation
are discussed with some considerable care in Section 4.2 prior to the
numerical construction in Section 4.3.

\subsection{The Recursion Relation}

As the unperturbed physical quantities vary like powers of the
cylindrical radius, the Mellin transform (e.g., Carrier, Krook \&
Pearson 1966) is a natural tool for simplifying the linear stability
equations~(\ref{eq:lindyn}). First, we collect the real-space
variables into a dimensionless four-component vector
\begin{equation}
(U_1,U_2,U_3,U_4)^T\equiv\left(\frac{\delta\Sigma}{\Sigma},
{\delta v_\theta \over c},{\delta\psi \over c^2} ,{
\delta v_R \over c}\right)^T\equiv\bU(R),
\end{equation}
with the corresponding Mellin transforms
$\bu(\alpha)\equiv(u_1,u_2,u_3,u_4)^T$ defined by
\begin{equation}
\bu(\alpha)\equiv \int\limits_0^\infty dR~R^{-1/2 -i\alpha}
\bU(R).
\label{Mellin}
\end{equation}
Both sides have the implicit $(\theta, t)$ dependence
$\exp(im\theta-i\omega t)$ of course.  This is not quite the usual
definition of a Mellin transform (Carrier, Krook \& Pearson
1966), but the factor of $R^{-1/2}$ in the integrand makes for a real
value of $\alpha_0$ as defined by eq.~(\ref{det}) below.

From the second and third of eqs.~(\ref{eq:lindyn}), eq.~(\ref{eq:Poisson}),
and from eq.~(\ref{eq:curleq}), we have the following system:
\begin{eqnarray}
{\omega\over ic} u_1(\alpha+i)
&=&-im\left[\Mach u_1(\alpha)+u_2(\alpha)\right]\nonumber\\
&&-(i\alpha-\half)u_4(\alpha),\nonumber\\
{\omega\over ic} u_2(\alpha+i)
&=&-im\left[\Mach u_2(\alpha)+ u_1(\alpha)-u_3(\alpha)\right]\nonumber\\
&& -\Mach u_4(\alpha),\nonumber\\
u_3(\alpha)&=& (\Mach^2+1)K(\alpha,m) u_1(\alpha),\nonumber\\
imu_4(\alpha)&=&(i\alpha+\half)u_2(\alpha) -\Mach u_1(\alpha).
\label{system}
\end{eqnarray}
Here, $K(\alpha,m)$ is the famous Kalnajs function (Snow 1952; Kalnajs
1971; Binney \& Tremaine 1987),
\begin{equation}
 	K(\alpha,m) = \half { {
	\Gamma \left[ \half \left( \half + m + i \alpha \right)  \right]
	\Gamma \left[ \half \left( \half + m - i \alpha \right)  \right]
	 } \over {
	\Gamma \left[ \half \left( {3 \over 2}+ m + i \alpha \right)  \right]
	\Gamma \left[ \half \left( {3 \over 2}+ m - i \alpha \right)  \right]
	} },
	\label{kgrav}
\end{equation}
which relates the Mellin transforms of the density and the potential
perturbations.  Note the distinction between $v_0^2\equiv 2\pi
G\Sigma_0R_0$ and $\vcirc^2=v_0^2\Mach^2/(\Mach^2+1)$.  Only the first
two of eqs.~(\ref{system}) couple different values of the Mellin
variable $\alpha$.  So, after elimination of $u_3$ and $u_4$ using the
last two equations, one has a system of the form
\begin{equation}
{\omega \over ic}\hu(\alpha+i)= \bM(\alpha)\cdot\hu(\alpha),
\label{recur}
\end{equation}
where $\hu=(u_1,u_2)^T$ and $\bM (\alpha)$ is the $2\times 2$ matrix
given by:
\begin{displaymath}
\bM =\left[\begin{array}{lr}{\displaystyle 
\frac{\Mach}{m}(\frac{i}{2}+\alpha -im^2)} 
&{\displaystyle \frac{\alpha^2+m^2+1/4}{im}} \\
\null & \null \\
{\displaystyle \frac{m^2 + \Mach^2 - m^2(\Mach^2+1)K}{im}} & 
{\displaystyle \frac{\Mach}{m}(\frac{i}{2} -\alpha -im^2)} 
\end{array}\right]
\label{matrix}
\end{displaymath}
It is helpful to view eq.~(\ref{recur}) as a recursion relation
relating Mellin transforms with arguments with the same real part, but
differing by imaginary increments. We shall refer to such an
arrangement as a ``ladder''.

For neutral modes, $\omega\to0$, there is no coupling between
different values of $\alpha$ in eqs.~(\ref{system}). Equivalently,
there is only one rung of the ladder. As realised by a number of
investigators (e.g., Zang 1976; Lemos \& Lynden-Bell 1993; Syer \&
Tremaine 1996; Evans \& Read 1998b), exact equiangular neutral modes
are possible.  In fact, a solution exists in the form of an exact
logarithmic spiral provided $\det \bM =0$, or
\begin{equation}
[1-(\Mach^2+1)K(\alpha,m)][\alpha^2 + m^2 + {1\over4}]-\Mach^2(m^2 - 2) =0.
\label{det}
\end{equation}
The roots of this equation are real. Let us call them $\pm\alpha_0$.
If we artificially set $K\to0$ thereby turning off the self-gravity,
$\alpha_0$ reduces to the quantity $\mu$ defined in
eq.~(\ref{eq:mudef}).

\subsection{Boundary Conditions}
Let us now consider two complementary, semi-infinite ladders differing
in the sign of the real part of the Mellin argument:
$\alpha=+\alpha_0+ni$ and $\alpha=-\alpha_0+ni$, where $n$ is
nonnegative integer that can become arbitrarily large. What are the
boundary conditions for the recurrence relation?  This is a delicate
matter and we will consider the bottom and top rungs of the ladders in
turn. In physical terms, these correspond to small-$R$ and large-$R$
respectively.
\subsubsection{The Top Rung}

At large radii, the WKBJ dispersion relation
\begin{equation}
\left(\omega-\frac{m\vcirc}{R}\right)^2=\frac{2\vcirc^2}{R^2}
-\frac{v_0^2}{R}k + c^2 k^2,
\end{equation}
indicates that self-gravity becomes negligible and the radial wavenumber
becomes constant, viz.,
\begin{equation}
k\sim\frac{\omega}{c}+\frac{(\Mach^2+1)/2~-m\Mach}{R}+O(R^{-2})
~~\mbox{if}~~R\gg\frac{\vcirc}{|\omega|}.
\end{equation}
The sign of $k$ has been chosen appropriately for the outgoing wave,
and the subdominant $O(R^{-1})$ terms have been explicitly retained as
they contribute logarithmically to the phase of the eigenfunction.  It
follows from equations~(\ref{eq:ydef}) -~(\ref{eq:yeqn}) that
\begin{equation}
\bU(R)\sim \exp\left(\frac{i\omega R}{c}\right)
R^{i[(\Mach^2+1)/2~-m\Mach]}\bU_\infty
\label{wynone}
\end{equation}
as $R\to\infty$, where $\bU_\infty$ is another constant vector.  This
can be compared with the eigenfunction in the non self-gravitating
case [defined by eqs~(\ref{eq:bigz})-(\ref{eq:comb})], where the
exponent is only $-im\Mach$. The term $(\Mach^2+1)/2$ in
eq.~(\ref{wynone}) stems ultimately from the self-gravity in the WKBJ
dispersion relation.  At the $n$th rung of the ladders, the Mellin
variable has had its imaginary part incremented $n$ times and has
become
\begin{equation}
\alphan = \pm \alpha_0 + in
\label{alphan}
\end{equation}
As $n\rightarrow \infty$, the Mellin transform (\ref{Mellin}) is
dominated by the large-$R$ behavior above, so that
\begin{eqnarray}
\hu(\alphan)&\sim&\hu_\infty\Gamma(s_n)
\left(\frac{ic}{\omega}\right)^{s_n},\nonumber \\
s_n &\equiv&
n+\half-i\left(\pm\alpha_0+m\Mach-\frac{\Mach^2+1}{2}\right)
\label{eq:largen}
\end{eqnarray}
The direction of the constant vector $\hu_\infty$ is determined by the
relations~(\ref{eq:ydef})-(\ref{eq:Ddef}). The second component of
$\hat u_{\infty}$ is $O(n^{-1})$ smaller than the first component
because $\delta v_\theta=y/R\sim
R^{-1}(\delta\Sigma/\Sigma)$ asymptotically.  As the overall
normalization of the eigenfunction is arbitrary, the simplest
approximation is
\begin{equation}
\hu_\infty= (1,0)^T.
\label{normal}
\end{equation}

In fact, though the WKBJ analysis gives good physical insight into
what is happening at the top rung of the ladders, the boundary
conditions~(\ref{eq:largen}) - (\ref{normal}) are insufficiently accurate
for numerical construction of the modes.  This compels us to develop
an asymptotic analysis to find the higher order corrections.  The
following paragraphs accomplish this task. They are primarily matters
of detail rather than principle. Readers interested mainly in the
latter can skip to the next sub-section without losing the thread of our
argument. 

Rather than deal with two coupled first-order difference equations, it
proves convenient to recast (\ref{recur}) as a single second-order
difference equation for which standard asymptotic techniques are
readily available in the literature (e.g., Bender \& Orszag 1978,
chap. 5).
Let us write the two-component vector $\hu(\alphan)$ as
\begin{equation}\label{eq:xydef}
\hu(\alphan) = \left[ \frac{ic}{\omega} \right ]^n (x_n,y_n). 
\end{equation}
Substitution into eq.~(\ref{recur}) yields a coupled pair of
first-order difference relations relating $(x_{n+1},y_{n+1})$
to $(x_n,y_n)$.
By eliminating the $y$s, one has a second-order difference
equation:
\begin{equation}\label{eq:threeterm}
x _{n+1} - \left( A_n + {B_n D_{n-1}\over B_{n-1}} \right)x_n
+ {B_n \Delta_{n-1} \over B_{n-1}} x_{n-1} =0.
\label{onediff}
\end{equation}
with
\begin{eqnarray}
A_n & = &  \frac{\Mach}{m}(\frac{i}{2}+\alphan -im^2)  \sim O(n)
\nonumber \\
B_n & = & \frac{\alphan^2+m^2+1/4}{im}  \sim O(n^2) \nonumber \\
C_n & = & \frac{m^2 + \Mach^2 - m^2(\Mach^2+1)K(\alphan,m)}{im}  \sim
O(1) \nonumber \\
D_n & = & \frac{\Mach}{m}(\frac{i}{2} -\alphan -im^2)  \sim O(n)
\label{onediffcoms}
\end{eqnarray}
This notation is used because the matrix $\bM_n = \bM(\alphan)$ which
takes us from the $n$th to $(n+1)$th rung of the ladder is just
\begin{equation}
\bM _n =\left[\begin{array}{lr}A_n &B_n \\
C_n & D_n 
\end{array}\right].
\end{equation}
It has determinant
\begin{equation}
\Delta_n = A_nD_n - B_nC_n
\label{deltan}
\end{equation}
For future use, let us also note that the second component $y_{n+1}$ can
be derived via
\begin{equation}\label{eq:ynp1}
y_{n+1} = (D_nx_{n+1} - \Delta_n x_n )/ B_n
\label{ydiffeqn}
\end{equation}
The coefficients of $x_{n+1}, x_n$ and $x_{n-1}$ in (\ref{onediff})
are $O(1), O(n)$ and $O(n^2)$ respectively. Therefore, this suggests
the ansatz (c.f. Bender \& Orszag 1978)
\begin{equation}
{x_{n+1}\over x_n} = P_1\alphan + P_0 + P_{-1}\alphan^{-1} +
                     P_{-2}\alphan^{-2} + \dots
\label{pansatz}
\end{equation}
where the $\{P_k\}$ are independent of $n$. By substituting this
ansatz into the difference equation (\ref{onediff}) and matching the
orders of the asymptotic expansion, it is possible to solve
recursively for the $\{P_k\}$.  A second-order linear recursion
relation should have two independent solutions, hence
there are two roots for the leading coefficient: $P_1=\mp i$.
Only the upper sign is consistent with the required
behaviour (\ref{eq:largen}), and the subsequent terms are then determined.
From the first two terms, we have
\begin{eqnarray}
{x_{n+1} \over x_n} & = &
n+\half-i\left(\pm\alpha_0+m\Mach-\frac{\Mach^2+1}{2}\right) 
+ O\left(\frac{1}{n}\right) \nonumber \\
{y_{n+1} \over x_{n+1}} & = &
-\frac{\Mach + i m}{n} + O\left( \frac{1}{n^2} \right),
\label{results}
\end{eqnarray}
where the second line has been derived with the aid of
(\ref{ydiffeqn}).  This suggests that as $n \rightarrow \infty$, the
solution on the $n$th rung approaches
\begin{equation}\label{eq:better}
\hu(\alphan) \sim
\Gamma(s_n)\left(\frac{ic}{\omega}\right)^{s_n}
\left(1, -\frac{\Mach + im}{n} \right)^T.
\end{equation}
This is recognised as a more accurate version of eqs~(\ref{eq:largen}) -
(\ref{normal}). 

From a computational point of view, it is best to carry out the
calculation (eased by computer algebra) to still higher order.  We
find that:
\begin{equation}
x_n  \sim  \Gamma(-i\alpha_n) (-i\alpha_n)^\nu\left[ 1
+ {if_1 \over \alphan} + O({1 \over \alphan^2}) \right] 
\label{largennew}
\end{equation}
with
\begin{eqnarray}
\nu & = & {1\over 2} - i\left(m\Mach-\frac{\Mach^2+1}{2}\right) \\
f_1 & = & {1\over 2} \Bigl[ (2-m^2)\Mach^2 - {3\over4}
     +m^2 + 2i (m\Mach -\Mach^2 -1) \nonumber \\
    &   & \qquad\qquad\qquad\qquad\qquad +2\nu(\nu + im\Mach) \Bigr] 
\end{eqnarray}
The second component $y_n$ can be found, for example, by use of
eq.~(\ref{ydiffeqn}). These are the boundary conditions that are
employed in our numerical implementation in Section 4.3 below.

%
%
%
%
%
%
%

\subsubsection{The Bottom Rung}
Now, let us see how the bottom rungs of these ladders
($\alpha=\pm\alpha_0$) correspond to the ingoing and outgoing waves
far inside the corotation radius. The Mellin transforms have poles at
$\alpha=\pm\alpha_0$ because there is an infinite range of $\ln R$
over which the eigenfunction is asymptotically equal to a sum of two
power laws.  In fact, this representation is exact for a neutral mode
($\omega=0$), where the Mellin transforms vanish on the higher rungs
of the ladder.  In the general case $\omega\ne 0$, one expects that
the growing modes are superpositions of the neutral modes at least for
$R\ll\ \vcirc /|\omega|$. In other words,
\begin{equation}
\bU(R)\sim \bU_+ R^{-1/2+i\alpha_0} +\bU_- R^{-1/2-i\alpha_0},
\label{asymp}
\end{equation}
where the coefficients $\bU_\pm$ are constant vectors.  Hence, if the
range of integration in (\ref{Mellin}) is divided into two intervals
$0<R\le R_0$ and $R_0\le R<\infty$, with $R_0\ll\vcirc/|\omega|$, then
the contribution from the first interval is approximately
\begin{displaymath}
\frac{\bU_+ R_0^{+i\alpha_0-i\alpha}}{i(\alpha_0-\alpha)}
+\frac{\bU_- R_0^{-i\alpha_0-i\alpha}}{i(-\alpha_0-\alpha)},
\end{displaymath}
provided $\mbox{Imag}(\alpha)>0$.  Guided by this, we expect that
$\hu(\alpha)$ is composed of both a regular part and a singular part
with simple poles:
\begin{equation}
\hu(\alpha)\to  \hu_{\rm reg}(\pm\alpha_0) + \frac{\hu_\pm}
{\alpha\mp\alpha_0}~~~\mbox{as}~~\alpha\to\pm\alpha_0 + i0^+.
\label{poles}
\end{equation}
Here, $\hu_\pm$ are two-component constant vectors linearly related to
$\bU_\pm$, while $\hu_{\rm reg}(\alpha)$ is a vectorial function that is
regular at the poles.  Furthermore, it must be true that
\begin{equation}
\bM(\alpha_0)\hu_+ = 0 = \bM(-\alpha_0)\hu_-,
\label{kernel}
\end{equation}
so that $\hu(\pm\alpha_0+i)$ is finite:
\begin{eqnarray}
-i{\omega \over c}\hu(\pm\alpha_0+i) & = &
\lim\limits_{\epsilon\to 0^+}\bM(\pm\alpha_0+i\epsilon)
\cdot\hu(\pm\alpha_0+i\epsilon) \nonumber \\
& = & \left.\frac{d\bM}{d\alpha} \right|_{\pm\alpha_0}\cdot\hu_\pm
 + \bM(\pm\alpha_0)\cdot \hu_{\rm reg}(\pm\alpha_0)\nonumber\\
\label{resid}
\end{eqnarray}
Owing to the conditions (\ref{kernel}), the directions of the vectors
$\hu_\pm$ are fixed.  Their complex normalization factors are related
by the requirement that the ingoing and outgoing parts of the
asymptotic form (\ref{asymp}) have equal moduli.  Therefore,
\begin{equation}
|\hu_+| = |\hu_-|.
\label{smalln}
\end{equation}
In other words, the residues of the poles at $\alpha=\pm\alpha_0$ are
related by the boundary conditions at small $R$.  Since the ingoing
and outgoing wave must carry equal energies, the moduli of these two
residues must be equal.  The relative phase of the two residues is
arbitrary in the absence of a condition on the phases of the ingoing
and outgoing waves.

Thus the large-$R$ boundary condition gives us two complex
constraints, or four real ones, at large $n$ (one complex constraint
for each of the two ladders). The small-$R$ boundary condition gives a
single real constraint.  Our degrees of freedom are the complex
residues of the poles at the bottom rung of each ladder (two complex
or four real parameters), plus the complex parameter $\omega$.  Hence
we have five real constraints and six real free parameters.  As in the
non self-gravitating case, {\it the problem is indeterminate unless we
impose an additional constraint}.  This could be the relative phase of
the residues at $\pm\alpha_0$, or else the real part of $\omega$.

At the risk of repetition, we now show that just as in the
non self-gravitating case of Section 3, so here also the ratio of moduli of
the ingoing and outgoing waves is determined by $\arg(\omega)$, while
the relative phase is determined by $|\omega|$.  On the one hand, we
have seen that the complex amplitudes of the outgoing and ingoing
waves at small $R$ are proportional to $\hu_+$ and $\hu_-$.  On the
other hand, the recursion relations
(\ref{eq:threeterm})-(\ref{eq:ynp1}) for the quantities $\{x_n,y_n\}$
defined by eq.~(\ref{eq:xydef}) are independent of $\omega$.  Therefore
the ratio of ingoing to outgoing complex amplitudes is
\begin{eqnarray}
\frac{\hu_+}{\hu_-}&\propto& \frac{\hu_n(\alpha_0+in)}{\hu_n(-\alpha_0+in)}
\nonumber \\
&\propto& \omega^{2i\alpha_0}=\exp\left[-2\alpha_0\arg(\omega)\right]
\cdot\exp\left[2i\alpha_0\ln|\omega|\right],
\end{eqnarray}
where the unwritten constants of proportionality are independent of
$\omega$ and $n$ as $n\to\infty$ [cf. eq.~(\ref{eq:better})].
Therefore all boundary conditions that neither absorb nor emit energy
produce the same ratio of real to imaginary parts for all
eigenfrequencies, but inasmuch as they fix different phase shifts at
the inner boundary, they produce different discrete values for the
moduli of the eigenfrequencies.  Also, successive eigenfrequencies are
always separated by $\pi/\alpha_0$ in the natural logarithm.

\subsection{Numerical Strategy and Results}

To construct the normal modes, we must adjust the complex parameter
$\omega$ so that all the boundary conditions are satisfied. For a
large value of $n$, the starting value of $\hu(\alphan)$ is fixed on
the two ladders using eq.~(\ref{largennew}). The recurrence
relation~(\ref{recur}) is successively applied to reduce the imaginary
part of the Mellin transform variable to unity.  This calls for the
evaluation of the Kalnajs function (\ref{kgrav}) for complex values of
$\alphan$ with large imaginary parts. The best way to proceed is to
use Stirling's formula to derive the asymptotic result
\begin{equation}
K(\alphan,m) \sim \pm {1\over \alphan}\left[ 1 + {1-4m^2 \over
8\alphan^2} \right]
\end{equation}
and then use the recurrence relation
\begin{equation}
K(\alphanm,m) = {m^2 + (\alphanm + {3\over2}i)^2  \over
                 m^2 + (\alphanm + {1\over2}i)^2 }
                  K(\alphan,m)
\end{equation}
to work down the ladders.  The final step to the bottom rung warrants
further discussion. From eq.~(\ref{resid}), the quantity
$\hu(\pm\alpha_0+i)$ contains contributions from both the singular
terms (which lie in the null eigenspace) and the non-singular terms
(which lie in its complement). Let $\bn$ be the null eigenvector so
that $\bM \cdot \bn =0$ and let $\bp$ be a projection operator that
annihilates the non-null eigenvector. Then the residues we seek on
each ladder are
\begin{equation}
\hu_\pm = \left[{\bp^T\cdot \hu(\pm \alpha_0 +i) \over \bp^T \cdot
\bM'(\pm \alpha_0)\cdot \bn}\right] \bn
\end{equation}
To satisfy the inner boundary condition, these residues must be equal
in modulus.  The entire numerical algorithm can be tested in the
non self-gravitating instance by the obvious stratagem of setting the
Kalnajs gravity factor $K(\alpha,m)$ to vanish. Here, the exact answer
is already known by virtue of the work in Section 3.

\begin{figure}
\psfig{figure=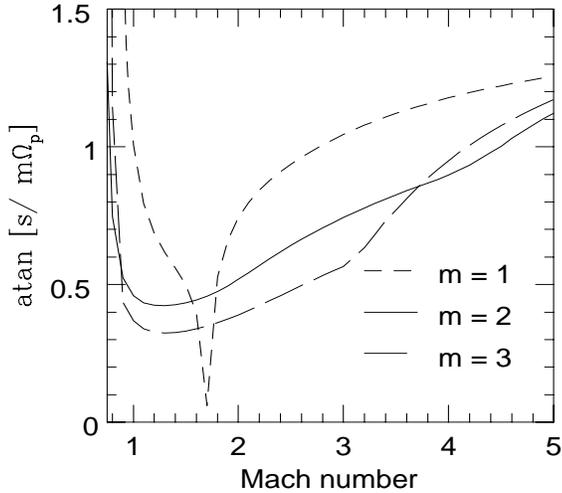,height=8.0truecm,width=8.4truecm}

\caption{Plot of the argument of the eigenfrequency (or arctangent of
the ratio of growth rate to pattern speed $\atan[ s/m\Omegap]$) as a
function of the Mach number $\Mach$ for self-gravitating modes.  The
diagram shows the dispersion relation for modes with azimuthal
wavenumber $m =1,2$ and 3.}
\end{figure}

Fig.~2 shows a graph of the ratio of growth rate to pattern speed for
self-gravitating modes in the Mestel disk. Now, growing modes are
possible provided the disk is cold enough. The condition for instability
can be worked out exactly as
\begin{equation}
\Mach^2 \ge {[m^2 +{1\over4}][1 - K(0,m)]\over
              m^2 - 2 + (m^2 +{1\over4})K(0,m)}
\end{equation}
where $K(0,m)$ is readily deduced from eq.~(\ref{kgrav}) as
\begin{equation}
 	K(0,m) = \half { {
	\Gamma^2 \left[ \half \left( \half + m \right)  \right]
	 } \over {
	\Gamma^2 \left[ \half \left( {3 \over 2}+ m \right)  \right]
	} }.
\end{equation}
So, for example, $m=1$ growing modes are possible only if $\Mach \ge
0.869$, and $m=2$ modes only if $\Mach \ge 0.733$. At a fixed
temperature or Mach number, the ratio of growth rate to pattern speed
is fixed. But, just as in the non self-gravitating case of Section
3.3, the magnitudes of these quantities are not fixed as a consequence
of the indeterminate nature of the phase shift at the origin.

\section{Conclusions}

This paper has resolved the paradox of the scale-free disks. By
itself, the linear stability problem is ill-posed. Normal modes do
exist, but the moduli of the eigenfrequencies depend on the
undetermined phase shift with which waves are reflected from the
centre. For any definite choice of this phase, there can be a discrete
set of growing modes for some disks.  On the other hand, the argument
of the eigenfrequencies---or equivalently the ratio of growth rate to
pattern speed---is independent of the phase shift.  This paper has
explicitly calculated such ratios with and without self-gravity.

The paradox of the scale-free disks is of interest for two
reasons. First, as Birkhoff (1960) has indicated, paradoxes are of
inherent pedagogical value. It sharpens the insight to track down the
flaw in a plausible physical argument that nonethless leads to an
apparent inconsistency.  The paradox of the scale-free disks is a
paradox of over-simplification.  In theoretical work, it is customary
to develop as simple a model as possible to describe objects or
phenomena. The power-law disks, in which all physical quantities scale
like powers, are attractive candidates for representing both galactic
and accretion disks. But, they are not sufficiently rich to act as
reasonable models for stability problems -- unless extra ingredients
such as reflecting boundaries or cut-outs are added. This is the
procedure that was followed in Zang (1976) and Evans \& Read (1998a,b)
to yield normal modes.  Second, the paradox draws attention to the
importance of inner boundary conditions in all models with central
cusps. The stability properties are crucially affected by the manner in
which impinging waves are reflected off the cusp. Therefore, the
correct boundary condition at the centre -- whether applied directly
in a linear stability analysis or indirectly in a computer simulation
-- needs careful thought.  Thinking physically, central cusps are
usually a consequence of black holes. Probably the most astronomically
relevant boundary condition is to allow the wave to reflect off a
central black hole.

\section*{Acknowledgments}
We thank the Isaac Newton Institute for hospitality during its program
on the Dynamics of Astrophysical Disks, when this work began. JG is
supported by NASA Astrophysical Theory Grant NAG5-2796, and NWE by the
Royal Society.

%


\begin{thebibliography}{99}

\bibitem[Abramowitz \& Stegun 1970]{AS} Abramowitz M., Stegun
I.A., 1970.  Handbook of Mathematical Functions, Dover, New York

\bibitem[Bender \& Orszag 1978]{BO} Bender C., Orszag S., 1978.
Advanced Mathematical Methods for Scientists and Engineers,
McGraw-Hill, New York

\bibitem[Binney \& Tremaine]{BT} Binney J., Tremaine S.,
1987. Galactic Dynamics, Princeton University Press, Princeton

\bibitem[Birkhoff]{Garrett}
Birkhoff G.D., 1960, Hydrodynamics: A Study in Logic, Fact and
Similitude, Princeton University Press, Princeton

\bibitem[Carrier, Krook \& Pearson]{CKP}
Carrier G.F, Krook M., Pearson C.E., 1966, Functions of a Complex
Variable: Theory and Technique, McGraw-Hill, New York

\bibitem[Drury L O'C 1980]{Luke}
Drury L.O'C., 1980, MNRAS, 193, 337

\bibitem[Evans \& Read 1998a]{ERa}
Evans N.W., Read J.C.A., 1998a, MNRAS, 300, 83

\bibitem[Evans \& Read 1998b]{ERb}
Evans N.W., Read J.C.A., 1998b, MNRAS, 300, 106

\bibitem[Kalnajs 1971]{Kalnajs71}
Kalnajs A., 1971, ApJ, 166, 275.

\bibitem[Lemos, Kalnajs \& Lynden-Bell 1991]{LKL}
Lemos J.P.S., Kalnajs A., Lynden-Bell D., 1991, ApJ, 375, 484

\bibitem[Lynden-Bell \& Lemos 1993]{LBL}
Lynden-Bell D., Lemos J.P.S., 1993, Equiangular modes of
power-law disks, unpublished.

\bibitem[Mestel 1963]{Mestel}
Mestel L., 1963, MNRAS, 126, 553

\bibitem[Narayan, Goldreich, \& Goodman (1987)]{NGG}
Narayan R., Goldreich P., Goodman J., 1987, MNRAS, 228, 1

\bibitem[Schmitz \& Ebert 1987]{SE}
Schmitz F., Ebert R., 1987, AA, 181, 41

\bibitem[Shu 1992]{Shu}
Shu F.H. 1992. The Physics of Astrophysics: Gas Dynamics,
University Science Books, Mill Valley, California

\bibitem[Snow 1952]{Snow}
Snow, C. 1952. Hypergeometric and Legendre Functions with Applications
to the Integral Equations of Potential Theory, National Bureau of
Standards, Washington D.C.

\bibitem[Syer \& Tremaine (1996)]{ST96}
Syer D., Tremaine S., 1996, MNRAS, 281, 925

\bibitem[Toomre 1964]{Toomre64}
Toomre A., 1964, ApJ, 139, 1217.

\bibitem[Zang 1976]{Zang76}
Zang T.A., 1976, Ph. D. thesis, Massachusetts Institute of Tecnology.

\end{thebibliography}
\end{document}